\documentclass[12pt]{article}
\usepackage[dvips]{graphicx}         
\usepackage{bm}                      

\begin{document}
\def\vc{\phi}
\def\tsurf{\sigma}
\def\rv{\bm{r}}

\title{Quark deconfinement transition in hyperonic matter}
 
\author{
Toshiki Maruyama,$^1$ 
Satoshi Chiba,$^1$ 
Hans-Josef Schulze,$^2$ \\
and Toshitaka Tatsumi$^3$}

\date{}

\maketitle 

\noindent{\it
$^{1}$ Advanced Science Research Center, Japan Atomic Energy Agency, 
       Tokai, Ibaraki 319-1195, Japan \\
$^{2}$ INFN Sezione di Catania, Via Santa Sofia 64, I-95123 Catania, Italy \\
$^{3}$ Department of Physics, Kyoto University, Kyoto 606-8502, Japan}

\begin{abstract}
We discuss the properties of the hadron-quark mixed phase in compact stars
using a realistic equation of state of hyperonic matter and the MIT bag model.
We find that the equation of state of the mixed phase is similar to that given by
the Maxwell construction, but that the mixed phase becomes mechanically
unstable if the surface tension of the interface between the two pure phases 
is strong enough.
The composition of the mixed phase is very different
from that of the Maxwell construction; in particular,
hyperons are completely suppressed.

\end{abstract}



It is commonly believed that hyperons appear in dense nuclear
matter at baryon densities above 2--3 times normal density,
in spite of some uncertainties about the nucleon-hyperon (NY) 
and hyperon-hyperon (YY) interactions.
Many theoretical studies have shown that the hadronic
equation of state (EOS) becomes very soft
once hyperons become components of the matter \cite{hyp,hypns}. 
As a major consequence, the maximum mass of neutron stars (NS) 
predicted using the hyperonic EOS
may remain below the current observational values of about 
1.5 solar masses \cite{pag}.

Some authors have suggested that this situation might be 
remedied by considering the yet unknown three-body forces (TBF) 
among hyperons and nucleons \cite{nis},
while other studies have shown that a quark deconfinement phase
transition in hyperonic matter renders the EOS sufficiently stiff again 
to allow NS masses consistent with current data \cite{qmns}.

However, the appearance of quark matter (QM) poses the problem of
an accurate theoretical description of the quark phase,
which is so far an open question,
and furthermore of the
details of the phase transition between hadronic and quark matter.
The purpose of this letter is the study of the latter problem, 
combining a Brueckner-Hartree-Fock (BHF) EOS of hyperonic hadronic matter 
with the standard phenomenological MIT model for the quark phase.
In the simplest scenario, the Maxwell construction (MC), a sharp transition
takes place between the two charge-neutral hadron and quark phases,
whereas the more general Gibbs (Glendenning) construction (GC) \cite{gle}
allows a mixed phase (MP) containing individually charged hadron 
and quark fractions with various geometrical structures.
However, in the latter case, electromagnetic and surface contributions
to the energy of the MP are usually neglected, but could have
important effects \cite{mix,vos,mixtat}.
The quantitative analysis of these corrections is the purpose of this letter.



Our theoretical framework for the hadronic matter
is the nonrelativistic BHF approach \cite{hypns,bhf}
based on microscopic
NN, NY, and YY potentials that are fitted to scattering phase shifts, 
where possible.
Nucleonic three-body forces (TBF) are included in order to (slightly) shift
the saturation point of purely nucleonic matter to the empirical value.
It has been demonstrated that the theoretical basis of the BHF method,
the hole-line expansion, is well founded:
the nuclear EOS can be calculated with good accuracy in the BHF two hole-line
approximation with the continuous choice for the single-particle potential,
since the results in this scheme are quite close to the full
convergent calculations which include also the three hole-line
contributions \cite{bhf,thl}.
Due to these facts, combined with the absence of adjustable parameters,
the BHF model is a reliable and well-controlled theoretical approach
for the study of dense baryonic matter.

The basic input quantities in the Bethe-Goldstone equation
are the NN, NY, and YY potentials.
In this work we use the Argonne $V_{18}$ NN potential \cite{v18} supplemented by
the Urbana UIX nucleonic TBF of Refs.~\cite{uix}
and the Nijmegen soft-core NSC89 NY potentials \cite{nsc89}
that are well adapted to the existing experimental NY scattering data
and also compatible with $\Lambda$ hypernuclear levels \cite{yamamoto,vprs01}.
With these potentials, the various $G$-matrices are evaluated
by solving numerically the Bethe-Goldstone equation. 
Then the total nonrelativistic hadronic energy density, $\epsilon_H$,
can be evaluated:
\begin{eqnarray}
 \epsilon_H \!&=& \!\!\!\!
 \sum_{i=n,p,\Lambda,\Sigma^-}
 \sum_{k<k_F^{(i)}}
 \left[ T_i(k) + {1\over2} U_i(k) \right] \:,
\label{e:eps}
\end{eqnarray}
with
$T_i(k) = m_i + {k^2\!/2m_i}$,
where the various single-particle potentials are given by
\begin{equation}
 U_i(k) =
 \sum_{i'=n,p,\Lambda,\Sigma^-} U_i^{(i')}(k)
\label{e:un}
\end{equation}
and are determined self-consistently from the $G$-matrices.


For the quark EOS, we use the MIT bag model with 
massless $u$ and $d$ quarks and massive $s$ quark with $m_s= 150$ MeV.
The quark matter energy density
can be expressed as a sum of the kinetic term
and the leading-order one-gluon-exchange term \cite{jaf,tama}
for the interaction energy
proportional to the QCD fine structure constant $\alpha_s$, 
\begin{eqnarray}
 \epsilon_Q &=& B + \sum_f \epsilon_f \:,
\\
 \epsilon_f(\rho_f) &=& {3m_f^4 \over 8\pi^2} \bigg[ 
 { \left(2x_f^3+x_f\right)\sqrt{1 + x_f^2}} - {\rm arsinh}\,x_f \bigg] 
\nonumber \\
 && - \alpha_s{m_f^4\over \pi^3} \bigg[ 
 x_f^4 - {3\over2}\Big( x_f \sqrt{1 + x_f^2} - {\rm arsinh}\,x_f \Big)^2 
 \bigg] \:,
\end{eqnarray}
where $m_f$ is the $f=u,d,s$ current quark mass,
$x_f = k_F^{(f)}\!/m_f$,
the baryon density of $f$ quarks is $\rho_f = {k_F^{(f)}}^3\!\!/3\pi^2$,
and the bag constant $B$ is the energy density difference between 
the perturbative vacuum and the true vacuum.

This is clearly an oversimplified model of QM, 
which will be used in this letter to study the generic qualitative
features of the hadron-quark phase transition in NS matter.
In future work we will compare in more detail
the quantitative results obtained using different,
more sophisticated, QM models.

\begin{figure}
\includegraphics[width=0.48\textwidth]{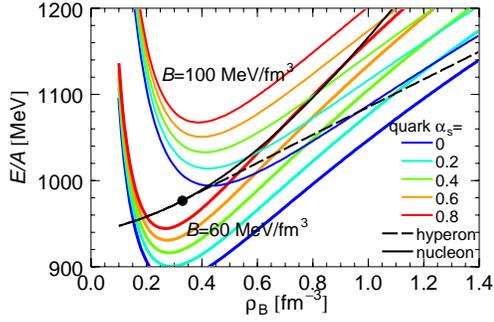}
\caption{
EOS of hadronic matter (black curves)
and of quark matter (colored curves)
with $B=60\;\rm MeV\!/fm^3$ (lower curves)
and $B=100\;\rm MeV\!/fm^3$ (upper curves) for several values of $\alpha_s$.
Hyperons appear at the dotted point in hadronic matter.}
\label{figParam}
\end{figure}

\begin{figure}
\includegraphics[width=0.48\textwidth]{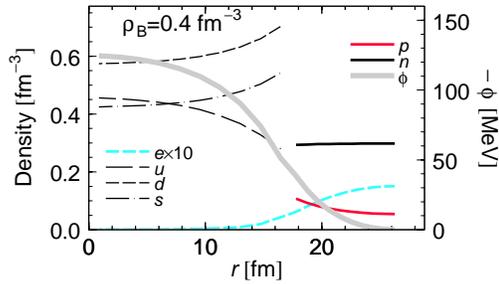}
\caption{
Density profiles 
and Coulomb potential $\vc$ 
within a 3D (quark droplet) Wigner-Seitz cell
of the MP at $\rho_B=0.4$ fm$^{-3}$.
The cell radius and the droplet radius are $R_W=26.7$ fm
and $R=17.3$ fm, respectively.
}
\label{figProf}
\end{figure}

\begin{figure}
\includegraphics[width=0.48\textwidth]{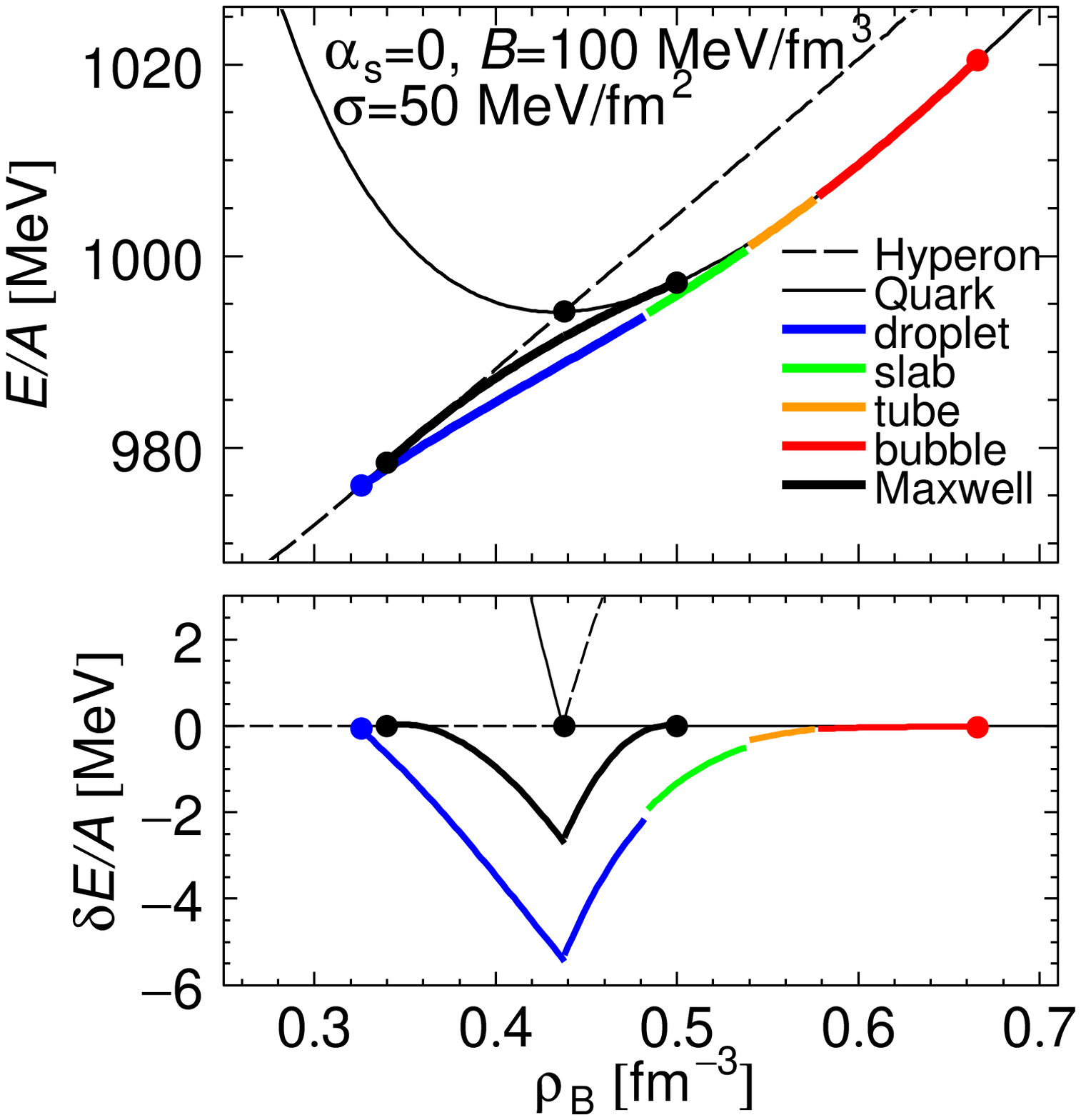}
\caption{
EOS of the MP (thick curves)
in comparison with pure hadron and quark phases (thin curves).
The upper panel shows the energy per baryon $E/A$ 
and the lower panel the energy difference between mixed and
hadron ($\rho_B<0.44$ fm$^{-3}$)
or quark ($\rho_B>0.44$ fm$^{-3}$) phases.
Different segments of the MP are chosen by minimizing the energy.
}
\label{figEOS}
\end{figure}

Figure~\ref{figParam} compares the hadronic BHF EOS 
and the quark matter EOS with 
different values of the parameters $B$ and $\alpha_s$
for beta-stable and charge-neutral matter.
One can see that the quark EOS approaches that of a relativistic
free gas ($E/A \sim \rho_B^{1/3}$) with increasing density, 
while the hyperonic EOS is always soft.
Consequently the quark deconfinement transition 
cannot occur at too high densities.
If we demand the quark and the hyperonic EOS to cross,
$\alpha_s$ should be small and $B$ slightly large, 
which gives a relatively low critical density.
Thus the appearance of hyperons is effectively suppressed 
due to a quark deconfinement transition.
In this letter we choose $\alpha_s=0$ and $B=100\;\rm MeV\!/fm^3$.


The numerical procedure to determine the EOS and the
geometrical structure of the MP is similar to that 
explained in detail in Refs.~\cite{mixtat}.
We employ a Wigner-Seitz approximation in which
the whole space is divided into equivalent Wigner-Seitz 
cells with a given geometrical symmetry,
sphere for three dimension (3D), cylinder for 2D, and slab for 1D.
A lump portion made of one phase is embedded in the other phase and thus 
the quark and hadron phases are separated in each cell.
A sharp boundary is assumed between the two phases and the surface energy
is taken into account in terms of a surface-tension parameter $\tsurf$.
We use the Thomas-Fermi approximation for the density profiles of
hadrons and quarks, 
while the Poisson equation for the Coulomb potential $\vc$ is explicitly solved.
The energy density of the mixed phase is thus written as
%
%
%
\begin{equation}
 \epsilon_M = {1\over {V_W}} \left[ 
 \int_{V_H} d^3 r \epsilon_H({\rv})+
 \int_{V_Q} d^3 r \epsilon_Q({\rv})+
 \int_{V_W} d^3 r \left( \epsilon_L({\rv}) + {(\nabla \vc({\rv}))^2\over 8\pi e^2} \right)
 + \tsurf S \right] \:,
\end{equation}
where the volume of the Wigner-Seitz cell $V_W$ is the sum of 
those of hadron and quark phases $V_H$ and $V_Q$,
and $S$ the quark-hadron interface area.
$\epsilon_L$ indicates the kinetic energy density of 
lepton (only the electron in this work).
The energy densities $\epsilon_H$, $\epsilon_Q$ and $\epsilon_L$ are 
$\rv$-dependent since they are functions of
local densities $\rho_a(\rv)$ ($a=n,p,\Lambda,\Sigma^-,u,d,s,e$). 
For a given density $\rho_B$, the optimum dimensionality of the cell,
the cell size $R_W$, the lump size $R$,
and the density profile of each component
are searched for to give the minimum energy density.
The structure of the MP changes from quark droplet to quark slab 
to hadron tube to hadron bubble with increasing baryon density.

The surface tension of the hadron-quark interface is poorly known, 
but some theoretical estimates based on the MIT bag model 
for strangelets \cite{jaf} and
lattice gauge simulations at finite temperature \cite{latt} suggest
a range of $\tsurf \approx 10$--$100\;\rm MeV\!/fm^2$.
We show results using $\tsurf=50\;\rm MeV\!/fm^2$ 
in the present letter, and discuss the effects of its variation.

\begin{figure}
\includegraphics[width=0.48\textwidth]{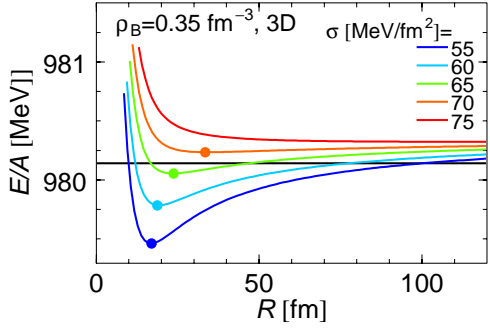}
\caption{
Droplet radius ($R$) dependence of the energy per baryon
for fixed baryon density $\rho_B=0.35$ fm$^{-3}$
and different surface tensions.
The quark volume fraction $(R/R_W)^3$ is fixed for each curve.
Dots on the curves show the local energy minima.
The black line shows the energy of the MC case.
}
\label{figRdep}
\end{figure}

\begin{figure}
\includegraphics[width=0.48\textwidth]{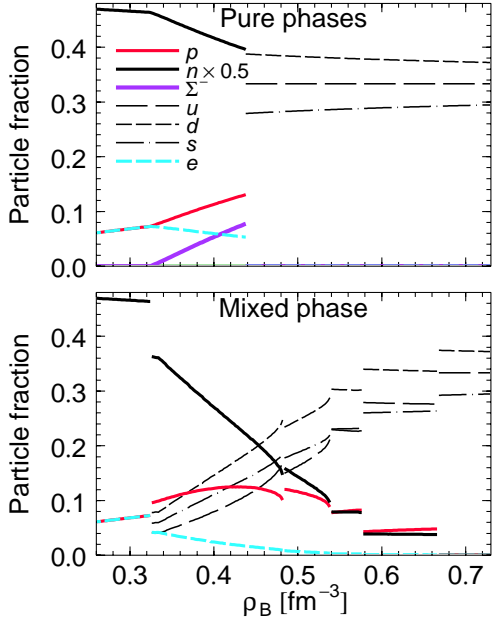}
\caption{
Particle fractions 
of quark and hadron species
in the pure hadron and quark phases (upper panel)
and in the MP (lower panel).
In the MC the phase transition occurs between the pure phases with
$\rho_B=0.34$ fm$^{-3}$ (hadrons) and 0.50 fm$^{-3}$ (quarks).
}
\label{figRatio}
\end{figure}

Figure~\ref{figProf} illustrates the outcome of this procedure,
showing the density profile in a 3D cell for $\rho_B=0.4$ fm$^{-3}$.
One can see the non-uniform density distribution of each particle species
together with the finite Coulomb potential.
The quark phase is negatively charged, so that 
$d$ and $s$ quarks are repelled to the phase boundary, 
while $u$ quarks gather at the center.
The protons in the hadron phase are attracted by the negatively charged 
quark phase, while the electrons are repelled.

Figure~\ref{figEOS} (upper panel) compares the resulting energy per baryon of 
the hadron-quark MP with that of the pure hadron and quark phases
over the relevant range of baryon density. 
The thick black curve indicates the case of the Maxwell construction,
while the colored line indicates the MP
in its various geometric realizations,
starting at $\rho_B=0.326$ fm$^{-3}$ with a quark droplet structure
and ending at $\rho_B=0.666$ fm$^{-3}$ with a hadron bubble structure.
The energy of the MP is only slightly lower than that of the MC, 
and the resultant EOS is similar to the MC one.
However, the structure and the composition of the MP
are very different from those of the MC case, as discussed later.

If one uses a smaller surface tension parameter $\tsurf$, 
the energy gets lower and the density range of the MP gets wider.
The limit of $\tsurf=0$ leads to a bulk application of 
the Gibbs conditions without the Coulomb and surface effects, i.e.,
the so-called Glendenning construction \cite{gle}.
On the other hand, using a larger value of $\tsurf$, 
the geometrical structures increase in size and
the EOS gets closer to that of the MC case. 
Above a limiting value of $\tsurf \approx 65\;\rm MeV\!/fm^2$ 
the structure of the MP becomes mechanically unstable \cite{vos}: 
for a fixed  volume fraction $(R/R_W)^3$
the optimal values of $R$ and $R_W$ go to infinity
and local charge neutrality is recovered in the MP, 
where the energy density equals that of the MC 
(see Fig.~\ref{figRdep}). 

This mechanical instability is due to the charge screening effect:
The optimal values of $R$ and $R_W$ are basically determined by the
balance between the Coulomb energy ($\sim R^2$ in the 3D case) 
and the surface energy ($\sim R^{-1}$).
However, if the charge screening is taken into account, the contribution of 
the screened Coulomb potential $\vc$ is strongly reduced 
when $R,R_W \rightarrow \infty$. 
A careful analysis by Voskresensky et al.~showed that the Coulomb 
energy changes its sign and behaves like
$R^{-1}$ as $R \rightarrow\infty$ due to the charge screening effect \cite{vos}. 
Thus the surface and the Coulomb energy give a local minimum below 
$\tsurf \approx 65\;\rm MeV\!/fm^2$,
which disappears when the surface energy
becomes greater than the Coulomb energy above 
$\tsurf \approx 65\;\rm MeV\!/fm^2$.
This is in contrast to the work of Heiselberg et al.~\cite{mix},
neglecting the charge screening effect,
where there is always a local energy minimum at finite $R$. The
importance of the charge screening effect has been also shown 
in the stability of strangelets \cite{hei}.

One notes in Fig.~\ref{figProf}
that no hyperons appear in the MP
although the mean baryon density $\rho_B=0.4\; \rm fm^{-3}$ 
is higher than the threshold density for hyperons 
in pure nucleon matter $\rho_B=0.34\; \rm fm^{-3}$ 
(see the black dot in Fig.~\ref{figParam}).
In the upper panel, the case of the pure hadron and quark phases
is shown for comparison.
One can see that the 
particle fractions 
are very different in both cases,
in particular a relevant hyperon fraction is only present in the 
hadronic part of the MC.

Thus we conclude that 
due to the relatively small magnitudes of the surface and Coulomb energies,
the EOS of the MP is similar to the MC one,
but the internal structure of the MP is very different.
In particular the role of hyperons is strongly reduced when we consider the
deconfinement transition in hyperonic matter.
Above a maximum value of the surface tension parameter, the MC is 
recovered as the physical one, however.
These results should be important for physical processes
like neutrino propagation and baryonic superfluidity,
besides the maximum mass problem, which will be studied
in an extended article.

This work is partially supported by the 
Grant-in-Aid for the 21st Century COE
``Center for the Diversity and Universality in Physics'' 
and the 
Grant-in-Aid for Scientific Research Fund 
of the Ministry of Education, Culture, Sports, Science and Technology of Japan
(13640282, 16540246). 



\begin{thebibliography}{99}

\bibitem{hyp}
%
 N.~K.~Glendenning,
 {\it Compact Stars: Nuclear Physics, Particle Physics and General Relativity},
 2nd ed. (Springer, Berlin, 2000);
 F.~Weber,
 {\it Pulsars as Astrophysical Laboratories for Nuclear and Particle Physics} 
 (IOP Publishing, Bristol, 1999).

\bibitem{hypns}
 M.~Baldo, G.~F.~Burgio, and H.-J.~Schulze,
 Phys.\ Rev.\ {\bf 58}, 3688 (1998);
 Phys.\ Rev.\ {\bf 61}, 055801 (2000);
 H.-J.~Schulze, A.~Polls, A.~Ramos, and I.~Vida\a~na,
 Phys.\ Rev.\ {\bf C73}, 058801 (2006).

\bibitem{pag}
For a recent review, see
 D.~Page and S.~Reddy,
 Annu.\ Rev.\ Nucl.\ Part.\ Sci.\ {\bf 56}, 327 (2006).

\bibitem{nis} 
 S.~Nishizaki, Y.~Yamamoto, and T.~Takatsuka,
 Prog.\ Theor.\ Phys.\ {\bf 105}, 607 (2001); 
 {\bf 108}, 703 (2002). 

\bibitem{qmns}
 G.~F.~Burgio, M.~Baldo, P.~K.~Sahu, and H.-J.~Schulze,
 Phys.\ Rev.\ {\bf C66}, 025802 (2002);
 M.~Baldo, M.~Buballa, G.~F.~Burgio, F.~Neumann, M.~Oertel, and H.-J.~Schulze,
 Phys.\ Lett.\ {\bf B562}, 153 (2003);
 C.~Maieron, M.~Baldo, G.~F.~Burgio, and H.-J.~Schulze,
 Phys.\ Rev.\ {\bf D70}, 043010 (2004).

\bibitem{gle}
 N.~K.~Glendenning,
 Phys.\ Rev.\ {\bf D46}, 1274 (1992);
 Phys.\ Rep.\ {\bf 342}, 393 (2001).

\bibitem{mix}
 H.~Heiselberg, C.~J.~Pethick, and E.~F.~Staubo,
 Phys.\ Rev.\ Lett.\ {\bf 70}, 1355 (1993);
 N.~K.~Glendenning and S.~Pei, 
 Phys.\ Rev.\ {\bf C52}, 2250 (1995);
 M.~B.~Christiansen and N.~K.~Glendenning,
 Phys.\ Rev.\ {\bf C56}, 2858 (1997);
 N.~K.~Glendenning,
 Phys.\ Rep.\ {\bf 342}, 393 (2001).

\bibitem{vos}
 D.~N.~Voskresensky, M.~Yasuhira, and T.~Tatsumi,
 Phys.\ Lett.\ {\bf B541}, 93 (2002);
 Nucl.\ Phys.\ {\bf A723}, 291 (2003);
 T.~Tatsumi, M.~Yasuhira, and D.~N.~Voskresensky,
 Nucl.\ Phys.\ {\bf A718}, 359 (2003).

\bibitem{mixtat}
 T.~Endo, T.~Maruyama, S.~Chiba, and T.~Tatsumi,
 Nucl.\ Phys.\ {\bf A749}, 333 (2005);
 T.~Endo, T.~Maruyama, S.~Chiba, and T.~Tatsumi,
 Prog.\ Theor.\ Phys.\ {\bf 115}, 337 (2006);
 T.~Maruyama, T.~Tatsumi, T.~Endo, and S.~Chiba,
 Recent Res.\ Devel.\ in Physics {\bf 7}, 1 (2006).

\bibitem{bhf}
 M.~Baldo, {\em Nuclear Methods and the Nuclear Equation of State}
 (World Scientific, Singapore, 1999).

\bibitem{thl}
 H.~Q.~Song, M.~Baldo, G.~Giansiracusa, and U.~Lombardo,
 Phys.\ Rev.\ Lett.\ {\bf 81}, 1584 (1998);
 M.~Baldo, G.~Giansiracusa, U.~Lombardo, and H.~Q.~Song,
 Phys.\ Lett.\ {\bf B473}, 1 (2000);
 M.~Baldo, A.~Fiasconaro, H.~Q.~Song, G.~Giansiracusa, and U.~Lombardo,
 Phys.\ Rev.\ {\bf C65}, 017303 (2002);
 R.~Sartor,
 Phys.\ Rev.\ {\bf C73}, 034307 (2006).

\bibitem{v18}
 R.~B.~Wiringa, V.~G.~J.~Stoks, and R.~Schiavilla,
 Phys.\ Rev.\ {\bf C51}, 38 (1995).

\bibitem{uix}
 B.~S.~Pudliner, V.~R.~Pandharipande, J.~Carlson, S.~C.~Pieper,
 and R.~B.~Wiringa,
 Phys.\ Rev.\ {\bf C56}, 1720 (1997);
 M.~Baldo, I.~Bombaci, and G.~F.~Burgio,
 Astron.\ Astroph.\ {\bf 328}, 274 (1997);
 M.~Baldo and L.~S.~Ferreira,
 Phys.\ Rev.\ {\bf C59}, 682 (1999);
 X.~R.~Zhou, G.~F.~Burgio, U.~Lombardo, H.-J.~Schulze, and W.~Zuo,
 Phys.\ Rev.\ {\bf C69}, 018801 (2004).

\bibitem{nsc89}
 P.~M.~M.~Maessen, Th.~A.~Rijken, and J.~J.~de~Swart,
 Phys.\ Rev.\ {\bf C40}, 2226 (1989).

\bibitem{yamamoto}
 Th.~A.~Rijken, V.~G.~J.~Stoks, and Y.~Yamamoto,
 Phys.\ Rev.\ {\bf C59}, 21 (1999).

\bibitem{vprs01}
 J.~Cugnon, A.~Lejeune, and H.-J.~Schulze,
 Phys.\ Rev.\ {\bf C62}, 064308 (2000);
 I.~Vida\~na, A.~Polls, A.~Ramos, and H.-J.~Schulze,
 Phys.\ Rev.\ {\bf 64}, 044301 (2001).

\bibitem{jaf}
 E.~Farhi and R.~L.~Jaffe,
 Phys.\ Rev.\ {\bf D30}, 2379 (1984);
 M.~S.~Berger and R.~L.~Jaffe,
 Phys.\ Rev.\ {\bf C35}, 213 (1987).

\bibitem{tama}
 R.~Tamagaki and T.~Tatsumi,
 Prog.\ Theor.\ Phys.\ Suppl.\ {\bf 112}, 277 (1993).

\bibitem{latt}
 K.~Kajantie, L.~K\"ark\"ainen, and K.~Rummukainen,
 Nucl.\ Phys.\ {\bf B357}, 693 (1991);
 S.~Huang, J.~Potvion, C.~Rebbi, and S.~Sanielevici,
 Phys.\ Rev.\ {\bf D42}, 2864 (1990); {\bf D43}, 2056 (1991).

\bibitem{hei}
 H.~Heiselberg, 
 Phys.\ Rev.\ {\bf D48}, 1418 (1993).\\
 P.~Jaikumar, S.~Reddy and A.~W.~Steiner, Phys.\ Rev.\ Lett.\ {\bf 96},
	040011 (2006);
 M.~G.~Alford, K.~Rajagopal, S.~Reddy and A.~W.~Steiner, Phys.\ Rev.\ {\bf
	D73}, 114016 (2006).
 


\end{thebibliography}
\end{document}